\documentclass{article}

\PassOptionsToPackage{numbers}{natbib}
\usepackage[final]{neurips_2025_ml4ps}
\usepackage{graphicx}
\usepackage{amsmath}
\usepackage{amssymb}
\usepackage{bbm}
\usepackage{url}
\usepackage{subcaption}
\usepackage{booktabs}
\title{Power law attention biases for molecular transformers}
\author{
    Jay Shen \\
    Department of Physics \\
    University of Chicago \\
    Chicago, IL 60637 \\
    \texttt{jshe@uchicago.edu} \\
    \And
    Oliver Tang \\
    Pritzker School of Molecular Engineering \\
    University of Chicago \\
    Chicago, IL 60637 \\
    \texttt{yifengt@uchicago.edu} \\
    \And
    Andrew Ferguson \\
    Pritzker School of Molecular Engineering \\
    University of Chicago \\
    Chicago, IL 60637 \\
    \texttt{andrewferguson@uchicago.edu} \\
}

\begin{document}

\maketitle

\begin{abstract}
    Transformers \cite{transformer} are the go-to architecture for most data modalities due to their scalability. While they have been applied extensively to molecular property prediction, they do not dominate the field as they do elsewhere \cite{transformers-mpp, esen}. One cause may be the lack of structural biases that effectively capture the relationships between atoms. Here, we investigate attention biases as a simple and natural way to encode structure. Motivated by physical power laws, we propose a family of low-complexity attention biases $b_{ij} = p \log|| \mathbf{r}_i - \mathbf{r}_j||$ which weigh attention probabilities according to interatomic distances. On the QM9 \cite{qm9} and SPICE \cite{spice} datasets, this approach outperforms positional encodings and graph attention while remaining competitive with more complex Gaussian kernel biases \cite{gaussian-kernel}. We also show that good attention biases can compensate for a complete ablation of scaled dot-product attention, suggesting a low-cost path toward interpretable molecular transformers. 
\end{abstract}

\section{Background}

\subsection{Vanilla scaled dot-product attention}

Most contemporary transformer flavors rely upon scaled dot-product attention, which computes the attention probabilities as: 
\begin{equation}\label{attention}
A_{ij} = \text{softmax}_j \Biggr[ \frac{\mathbf{q}_{i} \mathbf{k}_j^T}{\sqrt{d_k}} \Biggr]
\end{equation}
Here, $\mathbf{q}_{i} = \mathbf{x}_i\mathbf{W}_Q$ is the $i$th token's query vector, $\mathbf{k}_{j} = \mathbf{x}_j\mathbf{W}_K$ is the $j$th token's key vector, and $d_k$ is the query/key dimension. Implicitly, this formulation assumes that all information about both the tokens themselves and their structural relationships are contained in their embeddings $\mathbf{x}_i$. Most language and vision transformers incorporate that structural information by adding positional encodings conditioned on token position to the embeddings \cite{transformer, vit}. 

Positional encodings have been used in molecular transformers \cite{graphormer} with modest success. For example, the random walk positional encoding of Dwivedi et al. \cite{rwpe}, which tends to outperform others, samples random walks to capture information about an atom's bond neighborhood. 

\subsection{Attention biases}

Conceptually, positional encodings have a few disadvantages. For one, by their additive nature they are forced to associate with individual tokens rather than inter-token structures. In language or vision, this is not an issue, as tokens can be labeled with discrete, absolute positions. Molecules, however, are most naturally modeled as graphs or Euclidean point clouds, and neither modality admits such convenient indexes. 

An attractive alternative to positional encodings is the attention bias. Attention biases alter attention logits by adding some values $b_{ij}$ computed separately from the scaled dot-product:
\begin{equation}\label{biased-attention}
    A_{ij} = \text{softmax}_j \Biggr[ \frac{\mathbf{q}_{i} \mathbf{k}_j^T}{\sqrt{d_k}} + b_{ij} \Biggr]
\end{equation}
Attention biases naturally model pairwise relationships, and can easily be constructed given a measure of distance. For example, Liu et al. \cite{swin} construct attention biases for vision transformers based on the grid displacement between image patches. In the domain of molecular modeling, Luo et al. \cite{gaussian-kernel} learn Gaussian kernel functions that compute attention biases from interatomic distances. This approach achieves good performance on several molecular property prediction tasks. 

\subsection{Other structural biases}

Attention masking is another common structural bias which modulates information exchange between token pairs by setting the attention logit to $-\infty$. When applied to molecular models, it is often used to block interactions between non-bonded pairs: 
\begin{equation}\label{masked-attention}
    A_{ij} = \text{softmax}_j \Biggr[ \frac{\mathbf{q}_{i} \mathbf{k}_j^T}{\sqrt{d_k}} + M_{ij} \Biggr] \quad \text{where} \quad M_{ij} = 
    \begin{cases}
        0 & (i, j) \in \mathcal{E} \\
        -\infty & \text{otherwise}
    \end{cases}
\end{equation}
Attention masking is closely related to graph machine learning, especially graph attention kernels like GAT \cite{gat}. 

\section{Power law attention biases}

Motivated by physical power laws such as Coulomb's force, we propose the following attention bias:
\begin{equation}\label{power-law-bias}
    b_{ij} = p \log || \mathbf{r}_i - \mathbf{r}_j ||
\end{equation}
Because of the softmax operation, this bias term weights attention probabilities according to a power law of the interatomic distance: 
\begin{equation}
\begin{split}
    A_{ij} &= \text{softmax}_j \biggr [\frac{\mathbf{q}_{i} \mathbf{k}_j^T}{\sqrt{d_k}} + p \log || \mathbf{r}_i - \mathbf{r}_j || \biggr] \\
    &\propto \exp \biggr[ \frac{\mathbf{q}_{i} \mathbf{k}_j^T}{\sqrt{d_k}} + p \log || \mathbf{r}_i - \mathbf{r}_j || \biggr] \\
    &= || \mathbf{r}_i - \mathbf{r}_j ||^{p}\exp \biggr[ \frac{\mathbf{q}_{i} \mathbf{k}_j^T}{\sqrt{d_k}} \biggr]
\end{split}
\end{equation}
Here, $p$, which represents the power law exponent, is a parameter that can be held fixed, learned per layer, or learned per attention head for greater expressivity. It modulates attention probability separately from the query-key compatibility: this is conceptually similar to many interactions in physics. For example, Coulomb's law modulates the electric force using both particle charge and separation. 

One issue with this bias is that singularities appear when $i=j$ and $|| \mathbf{r}_i - \mathbf{r}_j || = 0$. We remedy this by simply masking out diagonal elements. We also experimented with adding a constant $\epsilon$ to the bias as well as learning separate self-interaction terms $b_{ii}$, but found no approach had any significant advantage over the others. 

\section{Experiments}

\subsection{Setup}

To compare the structural biases described above, we benchmark on the HOMO, LUMO, $U$, $G$, and $H$ energy targets from QM9 \cite{qm9}, as well as the $U_F$ energy target from SPICE \cite{spice}. Both QM9 and SPICE are quantum chemistry datasets generated using density functional theory. All regression targets were normalized, and samples were split into 8/1/1 train, validation, and test sets. QM9 was split by Murko scaffold while SPICE was split randomly. 

We standardized the parameter and compute budgets used across all tested models (Table \ref{tab:hyperparameters}). Models with attention biases only use them in the first four blocks, with the remaining blocks being vanilla transformers. 

\begin{table}[ht]
    \centering
    \caption{Standardized hyperparameters across all evaluated transformer models. Excluding structural bias modules, all models contained around 1.5M parameters. }
    \begin{tabular}{l c}
        \noalign{\vskip 1em}
        \toprule
        \noalign{\vskip 2pt}
        Embedding Dimension & 128 \\
        Number of Layers & 8 \\
        Attention Heads per Layer & 8 \\
        Dropout & 0.1 \\
        Training Epochs & 128 \\
        Batch Size & 64 \\
        Learning rate & 0.0001 \\
        Weight decay & 0.00001 \\
        Learning Rate Warmup & 1 epoch \\
        Learning Rate Schedule & On-plateau decay \\
        \bottomrule
    \end{tabular}
    \label{tab:hyperparameters}
\end{table}

\subsection{Results}

\begin{table}[ht]
    \centering
    \caption{
        Test mean absolute errors of transformer models with various structural biases when trained to predict energies from QM9 and SPICE. Reported values are the mean over an ensemble, with the standard deviations included, when significant, in parentheses. The number of additional parameters associated with each structural bias is reported, up to multiplicative and additive constants, in terms of the embedding dimension $E$, number of attention heads $H$, token dictionary size $T$, and number of transformer blocks with biased attention $N$. 
    }
    \begin{tabular}{lc c ccccc c c}
        \noalign{\vskip 1em}
        \toprule
        &&&\multicolumn{5}{c}{QM9} && \multicolumn{1}{c}{SPICE} \\
        \cline{4-8}\cline{10-10}
        \noalign{\vskip 2pt}
        Structural bias & \# Params && HOMO & LUMO & $U$ & $H$ & $G$ && $U_F$\\
        \hline
        \noalign{\vskip 2pt}
        
        Transformer \cite{transformer} & 0 && 0.34 & 0.74 & 24 (1) & 24 (1) & 24 (1) && 99 (8) \\
        Masked Attention & 0 && 0.12 & 0.14 & 29 (4) & 29 (4) & 30 (4) && - \\
        RWPE \cite{rwpe} & $kE$ && 0.14 & 0.20 & 42 (7) & 42 (7) & 42 (8) && - \\
        \noalign{\vskip 2pt}
        Attention bias & && & & & & && \\
        \cline{0-0}
        \noalign{\vskip 2pt}
        Gaussian kernel \cite{gaussian-kernel} & $kHT^2N$ && 0.08 & 0.10 & 21 (1) & 21 (1) & 21 (1) && - \\
        Power law $p=-1$ & 0 && 0.11 & 0.13 & 26 (1) & 26 (1) & 26 (1) && 7 (1) \\
        Power law $p \in \mathbb{R}$ & $HN$ && 0.12 & 0.14 & 25 (1) & 25 (1) & 25 (1) && 5 (1) \\
        Power law $p\in \mathbb{R}_-$ & $HN$ && 0.11 & 0.12 & 21 (1) & 21 (1) & 21 (1) && 5 (1) \\
        \bottomrule
    \end{tabular}
    \label{tab:maes}
\end{table}

\begin{figure}[ht]
    \centering
    \begin{subfigure}[t]{0.33\textwidth}
        \centering
        \includegraphics[width=\linewidth]{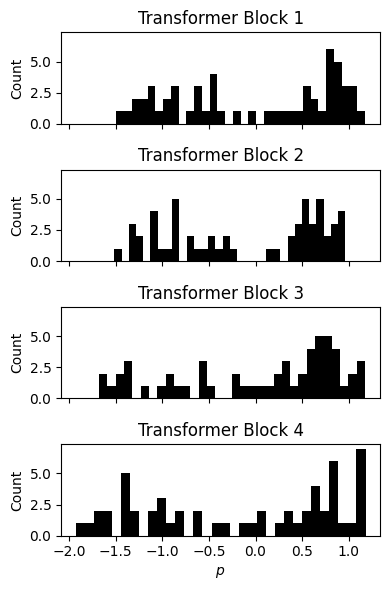}
        \caption{$p\in\mathbb{R}$}
    \end{subfigure}%
    \hspace{2.5em}
    \begin{subfigure}[t]{0.33\textwidth}
        \centering
        \includegraphics[width=\linewidth]{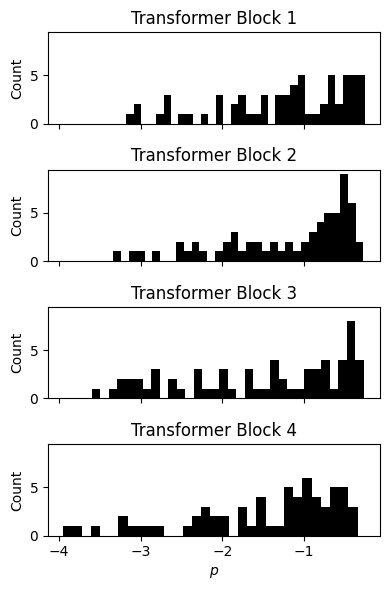}
        \caption{$p\in\mathbb{R}^-$}
    \end{subfigure}
    \caption{Histogram of learned power law exponents. Exponents are collected from an ensemble of 7 independently trained models. }
    \label{fig:exponents}
\end{figure}

The results of our experiments are described in \ref{tab:maes}. We find that attention biases consistently outperform the other classes of structural biases tested. Of the attention biases, Gaussian kernel biases are the most performant in general, but power law biases where $p\in\mathbb{R}^-$ match their accuracy on the $U$, $H$, and $G$ targets at much lower computational cost. 

It makes sense why the learned constrained power law $p \in \mathbb{R}^-$ is superior to power laws $p=-2$ and $p\in\mathbb{R}$, as it is more expressive than the former and, since physics is governed by inverse power laws, more principled than the latter. 

In Figure \ref{fig:exponents}, we report the distribution of exponents learned by ensembles of transformers with $p \in \mathbb{R}$ and $p \in \mathbb{R}$ power law attention biases. The exponents do not seem to tend toward any specific power laws and instead distribute randomly. This is to be expected, as neural networks are known to be highly uninterpretable. 

\subsection{Ablating scaled dot-product attention}

One interesting use of attention biases is as a substitute for scaled dot-product logits. Namely, instead of computing Equation (\ref{biased-attention}), we could compute attention patterns as:
\begin{equation}\label{fixed-attention}
    A_{ij} = \text{softmax}_j \big[ b_{ij} \big]
\end{equation}
This ``fixed'' attention completely decouples structure from token representations, unlike ``dynamic'' attention (Equation (\ref{biased-attention})) which allows embeddings to influence structural information flow via the dot-product logit. It this way, ``fixed'' attention is essentially message-passing on a fully connected graph. 

``Fixed'' attention has the immediate benefit of being cheaper to evaluate. It still scales quadratically as $\mathcal{O}(T^2)$, but does not scale with the embedding dimension $\mathcal{O}(ET^2)$ like ``dynamic'' attention. Here, $T$ refers to the token dictionary size and $E$ to the embedding dimension. ``Fixed'' attention also provides modest memory benefits, as the query and key weight matrices are no longer needed. 

\begin{table}[ht]
    \centering
    \caption{Ensemble mean percent change in MSE loss after ablating scaled dot-product attention}
    \begin{tabular}{l c ccccc}
        \noalign{\vskip 1em}
        \toprule
        Attention bias && HOMO & LUMO & $U$ & $H$ & $G$ \\
        \hline
        \noalign{\vskip 2pt}
        Gaussian kernel && +7.7\% & +7.4\% & -5.2\% & -5.6\% & -6.6\% \\
        Power law $p=-1$ && +13.8\% & +18.9\% & +3.3\% & +2.2\% & +4.3\% \\
        Power law $p\in \mathbb{R}$ && +7.2\% & +5.6\% & +2.7\% & +0.6\% & +1.9\% \\
        Power law $p \in \mathbb{R}_-$ && +3.8\% & +3.0\% & +16.0\% & +16.2\% & +16.9\% \\
        \bottomrule
    \end{tabular}
    \label{tab:ablating_attention}
\end{table}

\begin{figure}[ht]
    \centering
    \begin{subfigure}[t]{0.33\textwidth}
        \centering
        \includegraphics[width=\linewidth]{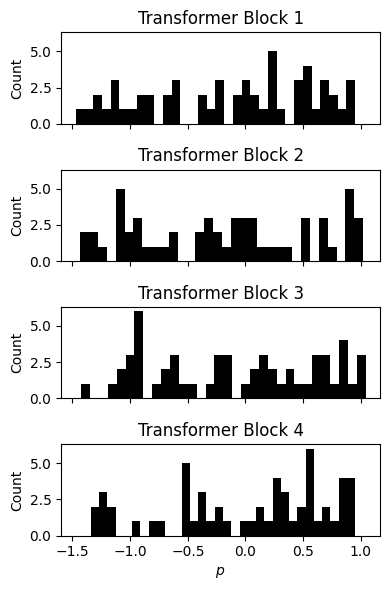}
        \caption{$p\in\mathbb{R}$}
    \end{subfigure}%
    \hspace{2.5em}
    \begin{subfigure}[t]{0.33\textwidth}
        \centering
        \includegraphics[width=\linewidth]{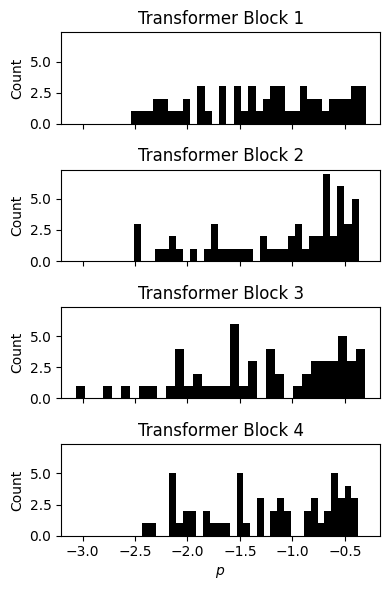}
        \caption{$p\in\mathbb{R}^-$}
    \end{subfigure}
    \caption{Histogram of learned power law exponents when dot-product attention is ablated. Exponents are collected from an ensemble of 7 independently trained models. }
    \label{fig:exponents_fixed}
\end{figure}

Table \ref{tab:ablating_attention} shows the percent change in loss when using ``fixed'' attention as opposed to ``dynamic'' attention. In general, the loss increases slightly, but in a few cases actually decreases, specifically for models with Gaussian kernel attention biases predicting the $U$, $H$, and $G$ targets. This may be because these energies are less dependent on complex molecule structure, and thus do not rely as heavily on information carried by ``dynamic'' attention. 

Figure \ref{fig:exponents_fixed} shows the distribution of exponents learned by ensembles of $p \in \mathbb{R}$ and $p \in \mathbb{R}$ power law bias models when scaled dot-product attention is ablated. The distribution of exponents is similarly uninterpretable like that of Figure \ref{fig:exponents}. 

We hypothesize that ``fixed'' attention layers may have use cases situated within larger models with ``dynamic'' attention layers. They can reduce compute and memory footprints while retaining or even enhancing accuracy. 

\section{Conclusions}

Here, we proposed a simple attention bias motivated by physics, showed its effectiveness at quantum chemical property prediction compared to baselines, and examined learned model representations. We also tested the hypothesis that attention biases can act as substitutes for scaled dot-product attention logits, and demonstrated that idea's feasibility. In addition to demonstrating empirical evidence in support of attention biases, we argued that they provide a more natural way to encode inter-atom structure in transformer models. 

The fundamental limitation of our study is scale: we experimented with one model size and two relatively small datasets. Future experimentation should scale up data, model size, and compute to clarify how our findings hold in different scenarios. For one, we proposed that replacing scaled dot-product attention may be useful when developing larger models—this hypothesis could be tested given greater scale. 

The code written for this work is available at \url{https://github.com/jshe2304/molecular_attention_bias}. 

\bibliography{references}

\end{document}